\documentstyle[aps,]{revtex}
\begin{document}
\title{Quantum Non-Demolition Bell State Measurement and N-party GHZ State
Preparation in Quantum Dot}
\author{Guo-Ping Guo\thanks{%
Electronic address: gpguo@ustc.edu.cn}, Hui Zhang, and Guang-Can Guo}
\address{Key Laboratory of Quantum Information, University of Science and\\
Technology\\
of China, CAS, Hefei 230026, People's Republic of China\bigskip \bigskip }
\maketitle

\begin{abstract}
By exploiting the fermionic qubit parity measurement, we present a scheme to
realize quantum non-demolition (QND) measurement of Bell-states and generate
n-party GHZ state in quantum dot. Compared with the original protocol, the
required electron transfer before and after parity measurement can be
nonadiabatic, which may speed up the operation speed and make the omitting
of spin-orbit interaction more reasonable. This may help us to construct
CNOT gate without highly precise control of coupling as the way of D.
Gottesman and I. L. Chuang.
\end{abstract}

\section{Introduction}

Since D. Loss and D. P. DiVicenzo proposed a quantum computation protocol
based on trapping spin electrons in semiconductor quantum dot (QD) in 1998,
the potential of QDs for implementing tasks in quantum information
processing (QIP) has been intensely studied both theoretically and
experimentally\cite{D. Loss and D. P. DiVincenzo,B. E Kane}. The spin of an
excess electron in the dot represents a promising qubit realization in such
systems and it can be more accessible and scalable compared with other
microscopic systems such as atoms or ions\cite{Jeroen Martijn Elzerman 1}.
In work\cite{D. P. DiVincenzo}, DiVincenzo put forward five criteria, which
must be all satisfied for any physical implementation of a quantum computer.
There is a detailed review about the experimental progress on the spin qubit
proposal using these five criteria\cite{J. M. Elzerman and etc. 9}. It has
been claimed that three criteria (well defined qubits, initialization and
readout) have been already achieved and future experiments may focus on
measuring the coherence time via the coherent manipulation of single spin
and the coherent coupling via the manipulation of spins in neighboring dots.

Although it has been shown\cite{E. Knill et al. Nature 409,M. A.
Nielsen,cnot} that partial measurement is sufficient for quantum computation
with photons, the extension of this paradigm to other systems is highly
desirable and challenging. In the so-called measurement-based quantum
computation, gates coupling qubits are no longer required. This eliminates
the need of highly precise control of the strength and pulsing between
qubits. Recently, H.-A. Engel and D. Loss presented a novel protocol of
electrons' spin parity measurement\cite{Hans-Andres Engel and Daniel Loss},
which explores the fact that resonant tunneling between the dots with
different Zeeman splitting is only possible when the spins are antiparallel.
By measuring the charge distributions between the two dots with a charge
detector, such as quantum point contact (QPC)\cite{J. M. Elzerman,M.
Kroutvar et al.,M. Field and et al.}, one can figure out that two electrons
initially loaded in one QD have either parallel or antiparallel spin
configuration without demolition. In addition, they suggested a construction
of CNOT gates in the way of C. W. J. Beenakker {\it et. al}\cite{cnot},
which requires two parity measurements, an ancillary qubit, a single-qubit
measurement, and the application of single-qubit operations depending on the
measurement outcomes. However, this CNOT construction requires that the
electron transferring between two dots should be adiabatic to keep the
electrons' spin states in each dot unchanged after a processing of loading
the two dots electrons together and then separating them\cite{guof}.

Consider two identical quantum dots 1 and 2 each with one electron initially
in state $(a\left| \uparrow \right\rangle _1+b\left| \downarrow
\right\rangle _1)$ and $(c\left| \uparrow \right\rangle _2+d\left|
\downarrow \right\rangle $\-$_2)$ . The subscripts 1 and 2 depict the
electrons in dot 1 and 2 respectively. The electron in dot 1 is transferred
into dot 2 by operation, which acts on electron charge degree of freedom and
keeps the electron spin state unchanged. Later, separate out one of the two
electrons in dot 2 back into dot 1. If these processes of transferring
electron into and out of dot 2 are adiabatic, the two electrons in dot 1 and
2 can respectively remain in spin states $(a\left| \uparrow \right\rangle
_1+b\left| \downarrow \right\rangle _1)$ and $(c\left| \uparrow
\right\rangle _2+d\left| \downarrow \right\rangle $\-$_2$ $)$ after these
processes\cite{Hans}. It is noted that the electron transfer between dots is
made by acting on electron charge degree of freedom, which preserves
electrons' spin state. If these electron transfer processes are
nonadiabatic, the electron in dot 1 can also be transferred into dot 2 while
its spin state is preserved. And the two electrons in dot 2 respectively in
spin states $(a\left| \uparrow \right\rangle _2+b\left| \downarrow
\right\rangle _2)$ and $(c\left| \uparrow \right\rangle _2+d\left|
\downarrow \right\rangle $\-$_2)$ will have equal probability to be
separated into dot 1 in these nonadiabatic transfer processes. Then there is
50\% probability that the two electrons in dot 1 and 2 will remain in spin
states 
\begin{equation}
(a\left| \uparrow \right\rangle _1+b\left| \downarrow \right\rangle
_1)(c\left| \uparrow \right\rangle _2+d\left| \downarrow \right\rangle _2)
\end{equation}
when one electron is transferred back into dot 1 with nonadiabatic
processes. There is also 50\% probability that the separated out electron
into dot 1 initially stays at dot 2 and in spin state $(c\left| \uparrow
\right\rangle _2+d\left| \downarrow \right\rangle $\-$_2)$. Then the two
electrons in dot 1 and 2 will changed into spin states 
\begin{equation}
(c\left| \uparrow \right\rangle _1+d\left| \downarrow \right\rangle
_1)(a\left| \uparrow \right\rangle _2+b\left| \downarrow \right\rangle _2).
\end{equation}
In this case, it seems as if the two electrons in dot 1 and 2 have exchanged
their spin states after all these processes. Although the nonadiabatic
electron transfer process can be performed faster than the adiabatic one, we
can not employ the C. W. J Beenakker {\it et. al } way\cite{cnot} to
construct CNOT gate if the two electrons after separating can be either
state (1) or state (2)\cite{guof,Hans}.

Here we propose a scheme to realize Bell-state QND measurement and prepare
n-party GHZ state with H.-A. Engel and D. Loss electron spin parity
measurement. Compared with the original protocol, the present scheme
emplores nonadiabatic electron transfer processes are emplored and there is
uncertainty in separating the two electrons with different spins from one
dot after the parity measurement. As single qubit rotations of electron in
individual quantum dot can be straightly achieved with radio-frequency
field, we can thus construct CNOT gate without highly precise controlled
coupling by exploring the electron parity measurement in the way of D.
Gottesman and I. L. Chuang \cite{Hans-Andres Engel and Daniel Loss,D.
Gottesman and I. L. Chuang}. \strut As the electron transferring between
dots, which is realized by acting on electron charge degree of freedom and
preserves the electron spin, does not need to be adiabatic, the operations
speeds may be enhanced and the neglecting of spin-orbital interaction may be
more reasonable.

\section{The Bell-state QND Measurement}

The idea of Bell-state QND measurement is firstly proposed by G.-P. Guo and
C.-F.Li\cite{G.-P. Guo and C.-F. Li}.If two qubits are initially in a Bell
state, the measurement can check which Bell state they are in without
destruction. And if the two qubits are not in Bell states, they can be
prepared in any Bell state. In this sense, the QND measurement can be used
as both a complete Bell-state analyzer and a Bell states generator. In the
original QND protocol, CNOT gates are employed, which challenges its
realization under the present experimental conditions. Now through the
electron spin parity measurement of H.-A. Engel and D. Loss, the Bell states
QND measurement can be implemented straightforwardly as shown in Fig 1, even
although nonadiabatic electron transferring is emplored.

In the first step, two electrons 1 and 2 are both loaded into dot A. This
electron transfer process can be made by acting on electron charge degree of
freedom and then accordingly can preserve electron spin state. In this
stage, the gate between the dots A and B is closed and the two electrons
stay in dot A. The two quantum point contact(QND) charge detectors $%
D_1(0)=1\ $\-and $D_2(0)=0$. In the second step, we open the gate for some
time $t$ (about 20ns\cite{Hans-Andres Engel and Daniel Loss}). If their
spins are antiparallel, the two electrons can tunnel to dot B. Then two
charge detectors will click as $D_1(t)=0\ $\-and $D_2(t)=1$. On the other
hand, if their spin are parallel, the two electrons will stay in dot A and
the detectors remains as $D_1(t)=1\ $\-and $D_2(t)=0$. In this way, we know
whether the two-electron spin state$\left| AA\right\rangle $ belongs to the
Hilbert space of states $\{\Phi ^{\pm }=\left( \left| \uparrow \uparrow
\right\rangle \pm \left| \downarrow \downarrow \right\rangle \right) /\sqrt{2%
}\}$ or $\{\Psi ^{\pm }=\left( \left| \uparrow \downarrow \right\rangle \pm
\left| \downarrow \uparrow \right\rangle \right) /\sqrt{2}$\-$\}$. This is
just the H.-A. Engel and D. Loss' original partial Bell-state measurement.

To realize Bell-state QND measurement, we proceed to separate the two
electrons into two dots with some static field acting on electron charge
degree of freedom. As the two electrons are separated in two dots,
radio-frequency fields can address individual electron spin. We can preform
single qubit operations and respectively rotate the two electrons spins as $%
\left| \uparrow \right\rangle \Rightarrow (\left| \uparrow \right\rangle
+\left| \downarrow \right\rangle )/\sqrt{2}$ , $\left| \downarrow
\right\rangle \Rightarrow (\left| \uparrow \right\rangle -\left| \downarrow
\right\rangle )/\sqrt{2}$. Accordingly, these Hadamard operations will
rotate the two-electron state as\cite{G.-P. Guo and C.-F. Li}:

\begin{equation}
\hat{H}_1\hat{H}_2\left( 
\begin{array}{c}
\Phi ^{+} \\ 
\Phi ^{-} \\ 
\Psi ^{+} \\ 
\Psi ^{-}
\end{array}
\right) =\left( 
\begin{array}{c}
\Phi ^{+} \\ 
\Psi ^{+} \\ 
\Phi ^{-} \\ 
-\Psi ^{-}
\end{array}
\right) .
\end{equation}
Then we reload the two electrons into one dot (for example dot A ), and
re-open the gate for some time $t.$ We observe the two detectors $D_1(2t)$
and $D_2(2t)$. Combined with $D_1(t)$ and $D_2(t)$, we can determine exactly
which Bell state the two electrons initially belong to just as shown in
Table 1.

\[
\text{Table 1: The states of the two detectors corresponding to each Bell
state} 
\]
\[
\begin{tabular}{|lllll|}
\hline
\multicolumn{1}{|l|}{} & \multicolumn{1}{l|}{$\Psi ^{+}$} & 
\multicolumn{1}{l|}{$\Psi ^{-}$} & \multicolumn{1}{l|}{$\Phi ^{+}$} & $\Phi
^{-}$ \\ \hline
\multicolumn{1}{|l|}{$\left| D_1(t)D_2(t)\right\rangle $} & 
\multicolumn{1}{l|}{$\left| 01\right\rangle $} & \multicolumn{1}{l|}{$\left|
01\right\rangle $} & \multicolumn{1}{l|}{$\left| 10\right\rangle $} & $%
\left| 10\right\rangle $ \\ \hline
\multicolumn{1}{|l|}{$\left| D_1(2t)D_2(2t)\right\rangle $} & 
\multicolumn{1}{l|}{$\left| 01\right\rangle $} & \multicolumn{1}{l|}{$\left|
10\right\rangle $} & \multicolumn{1}{l|}{$\left| 10\right\rangle $} & $%
\left| 01\right\rangle $ \\ \hline
\end{tabular}
\]
Lastly, we separate the two electrons and irradiate them to perform Hadamard
operations as the above steps. In this way, the final output state will
recover to the initial state if the two electrons is originally in a Bell
state and a complete Bell-state measurement is implemented without
destruction.

As all operations are identical to the two electrons, we need not know which
electron (with different spin state) is separated out from dot B in this
Bell-state QND measurement. Distinguished with the original protocol, all
the electron transfer processes between dots can thus be non-adiabatic and
directly realized by acting on electron charge degree of freedom.

Obviously, if the two electrons are initially in an arbitrary state $\Phi
_{12}=a\Phi ^{+}+b\Phi ^{-}+c\Psi ^{+}+d\Psi ^{-},$where $a,b,c,d\in C$, the
above measurement process will project them into one of the Bell states: 
\begin{eqnarray}
&&(a\Phi ^{+}+b\Phi ^{-}+c\Psi ^{+}+d\Psi ^{-})\otimes \left|
D_1(t)D_2(t)\right\rangle \otimes \left| D_1(2t)D_2(2t)\right\rangle \\
&\rightarrow &a\Phi ^{+}|10\rangle \otimes |10\rangle +b\Phi ^{-}|10\rangle
\otimes |01\rangle +c\Psi ^{+}|01\rangle \otimes |01\rangle +d\Psi
^{-}|01\rangle \otimes |10\rangle .  \nonumber
\end{eqnarray}
For example, we have a probability of $\left| d\right| ^2$ to project the
two electrons into state $\Psi ^{-}$. With some single qubit rotations, we
can thus get any other Bell-state. In this sense, the above measurement can
be also regarded as spin-electron Bell states generation protocol, which
theoretically has unit efficiency.

Actually, there is no need for two charge detectors, as the signal from D1
is always anti-correlated with the signal from D2. Therefore a single charge
detector is sufficient. This would simplify the experimental setup.

\section{Preparation of the n-particle GHZ state}

After the demonstration of two electrons full Bell states measurements, we
now discuss the preparation of the prior required GHZ entanglement states
for the CNOT gate construction\cite{D. Gottesman and I. L. Chuang}. We
assume that we have gotten two electrons $i,$ $j$ in the state $\Phi
_{ij}^{+}=\left( \left| \uparrow \uparrow \right\rangle +\left| \downarrow
\downarrow \right\rangle \right) /\sqrt{2}$ in dot B from the above
Bell-state QND measurement. Then we can separate one electron from dot B to
dot A by static fields, which preserve their total spin-state $\Phi ^{+}$%
\cite{W. G. van der Wiel}. Adjusting the bias voltage and the gate voltage
between dot B and dot C, we can load another electron $k$ from dot C to dot
B as shown in Fig.2. The initial state of the electron in dot C is $\Psi _k=%
\frac 1{\sqrt{2}}\left( \left| \uparrow \right\rangle _k+\left| \downarrow
\right\rangle _k\right) $\-.\- Thus the three-electron state can be written
as (only the electron spin wavefunctions are shown):

\begin{eqnarray}
\Psi _{ijk} &=&\Phi _{ij}^{+}\otimes \Psi _k=\frac 1{\sqrt{2}}\left( \left|
\uparrow \uparrow \right\rangle _{ij}+\left| \downarrow \downarrow
\right\rangle _{ij}\right) \otimes \frac 1{\sqrt{2}}\left( \left| \uparrow
\right\rangle _k+\left| \downarrow \right\rangle _k\right)  \nonumber \\
&=&\frac 12(\left| \uparrow \right\rangle _i\left| \uparrow \uparrow
\right\rangle _{jk}+\left| \downarrow \right\rangle _i\left| \downarrow
\downarrow \right\rangle _{jk}+\left| \uparrow \right\rangle _i\left|
\uparrow \downarrow \right\rangle _{jk}+\left| \downarrow \right\rangle
_i\left| \downarrow \uparrow \right\rangle _{jk}).
\end{eqnarray}
The two electrons (electron 2 and 3) in dot B have $1/2$ probabilities to be
measured in parallel spins to get state $(\left| \uparrow \right\rangle
_i\left| \uparrow \uparrow \right\rangle _{jk}+\left| \downarrow
\right\rangle _i\left| \downarrow \downarrow \right\rangle _{jk})/\sqrt{2}$.
In this case, when we separate one electron from dot B to dot C, the three
electrons ( $i,j,k$ named the three electrons respectively in dot A, B and
C) are in the state $(\left| \uparrow \right\rangle _i\left| \uparrow
\right\rangle _j\left| \uparrow \right\rangle _k+\left| \downarrow
\right\rangle _i\left| \downarrow \right\rangle _j\left| \downarrow
\right\rangle _k)/\sqrt{2},$ which can be rotated into any other GHZ state
with single qubit operation. Without doubt, the two electrons $j,k$ in dot B
have $1/2$ probabilities to be measured in anti-parallel spins to get state $%
(\left| \uparrow \right\rangle _i\left| \uparrow \downarrow \right\rangle
_{jk}+\left| \downarrow \right\rangle _i\left| \downarrow \uparrow
\right\rangle _{jk})/\sqrt{2}$. In view of the indistinguishability of two
electrons $j$ and $k$ in dot B and the uncertainty in separating them, we
then have equal probability to get the state $(\left| \uparrow \right\rangle
_i\left| \uparrow \right\rangle _j\left| \downarrow \right\rangle _k+\left|
\downarrow \right\rangle _i\left| \downarrow \right\rangle _j\left| \uparrow
\right\rangle _k)/\sqrt{2}$ and $(\left| \uparrow \right\rangle _i\left|
\downarrow \right\rangle _j\left| \uparrow \right\rangle _k+\left|
\downarrow \right\rangle _i\left| \uparrow \right\rangle _j\left| \downarrow
\right\rangle _k)/\sqrt{2}$, when we separate one electron from dot B to dot
C. In this case, we can then reload the two electrons $i$ and $j$ into one
dot and measure their parity again. After separating them, we then again
have $1/2$ probabilities to get the three electrons GHZ state $(\left|
\uparrow \right\rangle _i\left| \uparrow \right\rangle _j\left| \downarrow
\right\rangle _k+\left| \downarrow \right\rangle _i\left| \downarrow
\right\rangle _j\left| \uparrow \right\rangle _k)/\sqrt{2}$. Two electrons $%
i $ and $j$ also have $1/2$ probabilities to be measured in antiparallel
spins to get state $(\left| \uparrow \downarrow \right\rangle _{ij}\left|
\uparrow \right\rangle _k+\left| \downarrow \uparrow \right\rangle
_{ij}\left| \downarrow \right\rangle _k)/\sqrt{2}$. In this case, when we
separate electrons $i,j$, we can have again equal probability to get the
state $(\left| \uparrow \right\rangle _i\left| \downarrow \right\rangle
_j\left| \uparrow \right\rangle _k+\left| \downarrow \right\rangle _i\left|
\uparrow \right\rangle _j\left| \downarrow \right\rangle _k)/\sqrt{2}$ and $%
(\left| \downarrow \right\rangle _i\left| \uparrow \right\rangle _j\left|
\uparrow \right\rangle _k+\left| \uparrow \right\rangle _i\left| \downarrow
\right\rangle _j\left| \downarrow \right\rangle _k)/\sqrt{2}$. If necessary,
we can again measure the spin parity of electron $j$ and $k$. In the process
of repeating comparing the three electrons' spin, we can have a success
probability $p$ of $1-\frac 1{2^m}$ to get the electron GHZ state, where $m$
represents the times of comparing the electrons spins. When $m$ is large
enough, $p\rightarrow 1$ and we can almost prepare three electrons GHZ
state. Of course, if we only compare the spins of the two electrons $j,k$
initially in dot B one time, we have $1/2$ probabilities to get the GHZ
state $(\left| \uparrow \right\rangle _i\left| \uparrow \right\rangle
_j\left| \uparrow \right\rangle _k+\left| \downarrow \right\rangle _i\left|
\downarrow \right\rangle _j\left| \downarrow \right\rangle _k)/\sqrt{2}.$

\smallskip Obviously, the above process can be directly performed on the
case of n-electron GHZ states preparation ($n>3$) by comparing the spins of
the electrons between two dots in turn. In this way, the generation
efficiency is $p^{n-2},$where $p$ is the success probability of get three
electron GHZ states. In fact, the above GHZ state generation procedure is
very similar to the idea of S. Bose{\it \ et. al.\cite{swapping}, }which
shows that{\it \ }the entangled states involving higher number of particles
can be generated from entangled states involving lower number of particles
by employing the same procedure as entanglement swaps. The basic ingredient
of the original paper\cite{Guo,swapping} is a Bell state measuring device
and some L-particle (particle of lower number ($n=3$) ) entanglement states.
However, it can be proved that the required lower numbers of particles
entanglement states can be just 2-particle entangled states with the present
non-demolition partial Bell state measurement. Firstly, we prepare two
copies of entangled states of $n/2$ ( for example, assume $n$ is even)
electrons in the form $\Psi _{n/2}=(\left| \uparrow \uparrow \uparrow
...\uparrow \right\rangle _{123...\frac n2}+\left| \downarrow \downarrow
\downarrow ...\downarrow \right\rangle _{123...\frac n2}$\-$)/\sqrt{2}$.
Secondly, we coherently draw one electron from each of the two copies into a
dot. By checking these two electrons spins (if it is parallel), we can get
the entangled state of $n$ electrons $\Psi _n=(\left| \uparrow \uparrow
\uparrow ...\uparrow \right\rangle _{123...n}+\left| \downarrow \downarrow
\downarrow ...\downarrow \right\rangle _{123...n}$\-$)/\sqrt{2}$ with 1/2
probability. As the success probability of each time parity checking is 1/2,
this $n$ ($n\geq 2$) electrons preparation protocol has an efficiency of $%
p^{n/2-1}$ ($n$ is even) or $p^{^{(n+1)/2-1}}$ (n is odd). As the $n$
electrons in $\Psi _n$ are respectively in $n$ quantum dots, we can rotate
it into any other $n$-particle GHZ entangled states with single-qubit
operations.

\section{Conclusion}

\smallskip Here we propose a protocol to realize the QND measurement of
Bell-state, prepare $n$-electron GHZ states and then construct CNOT\ gate by
employing the novel partial Bell-state measurement of Fermionic qubits in
the article\cite{Hans-Andres Engel and Daniel Loss}. Distinguished with the
previous protocol\cite{Hans-Andres Engel and Daniel Loss}, the electron
transfer processes before and after the spin parity measurement can
nonadiabatic, which may make the omitting of the spin-orbit interaction
effects in electron transport more sensible and speed up the computation
operations. Furthermore, the present modified protocol has similar
robustness to the experimental noises, such as the effect of extra phases
from the inhomogeneous Zeeman splitting, finite J and different tunnel
couplings of singlet and triplet. It is noted that the precision of the
charge detectors or the fidelity of QPC measurements, which greatly
influence the success of the electrons spin parity checking and the present
protocols, have been recently analyzed in detail\cite{J. M. Elzerman}.

{\bf Acknowledgments}

This work is funded by the National Fundamental Research Program
(2001CB309300), National Nature Science Foundation of China (10304017), the
Innovation Funds from Chinese Academy of Sciences.

\-

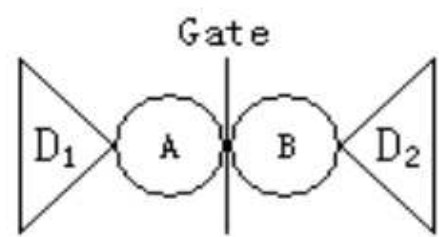
{\bf Fig. 1.} Schematic picture of the Bell-state analyzer, which includes
two coupled quantum dots (circle A and B), two quantum point contact charge
detectors (triangle D$_1$\- and D$_2$) and a gate (solid vertical line). Dot
A and B are assumed to have different Zeeman splitting, and the individual
tunneling events can be efficiently identified with a time resolved
measurement\cite{J. M. Elzerman,H. -A. Engel et al.,W. Lu and Z. Ji and L.
Pfeiffer and K. W. West A. J. Rimberg,R. Schleser et al.}. The gate switches
on and off the coupling between dot A and B. Here we consider only the case
that the two electrons are both in dot A or in dot B $\left| AA\right\rangle 
$\- and $\left| BB\right\rangle $, as the coupling to the state $\left|
AB\right\rangle $\- is small\cite{Hans-Andres Engel and Daniel Loss}. The
two electrons tunneling between states $\left| AA\right\rangle $\- and $%
\left| BB\right\rangle $ is only resonant, when they have antiparallel spins
and in the Hilbert space $\{\Psi ^{\pm }=(\left| \uparrow \downarrow
\right\rangle \pm \left| \downarrow \uparrow \right\rangle $\-$)/\sqrt{2}\}$%
. If they are in space $\{$ $\Phi ^{\pm }=(\left| \uparrow \uparrow
\right\rangle $\- $\pm \left| \downarrow \downarrow \right\rangle $\-$)/%
\sqrt{2}\}$, these two electrons will remain on the initial dot. This
requirement is the key principle of electrons spin parity measurements. If
the two electrons are both in the dot A (or in the dot B), $D_1=1\ $\-and $%
D_2=0\ $(or $D_2=1\ $\-and $D_1=0\ $).

\strut {\bf Fig. 2.} Schematic picture for preparation of the $3$-electron
GHZ states. Initially, dot A, B and C each has one electron. The electron
transfer between quantum dots is required to preserve electron spin state,
which can be achieved by static electrical fields acting on the electron
charge degree of freedom. Spin parity measurement of two electrons is made
in dot B.

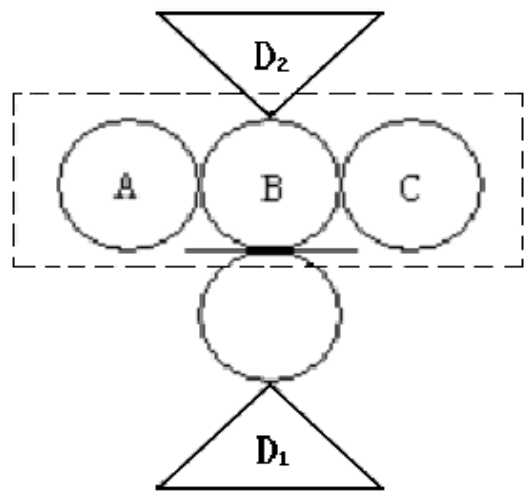
\end{document}